\documentclass[aps,prl,reprint,nofootinbib]{revtex4-1}

\usepackage{amsmath,amssymb,amsthm}
\usepackage{bbm}
\usepackage{bm}
\usepackage{graphicx}
\usepackage{hyperref}
\usepackage{mathrsfs}

\theoremstyle{definition}

\theoremstyle{remark}
\newtheorem*{rem*}{Remark}
\newtheorem*{note*}{Note}

\usepackage{xcolor}

\def\CL {{\cal L}}
\def\CN {{\cal N}}

\def\CK {{\cal K}}

\def\CG {{\cal G}}
\def\CR {{\cal R}}
\def\CT {{\cal T}}
\def\CQ {{\cal Q}}
\def\CO {{\cal O}}
\def\CI {{\cal I}}

\def\a{{\alpha}}

\def\sig{{\sigma}}

\def\p {{\partial}}
\def\be{\begin{equation}}
\def\ee{\end{equation}}
\def\bea{\begin{eqnarray}}
\def\eea{\end{eqnarray}}
\def\bes{\begin{subequations}}
\def\ees{\end{subequations}}

\def\oh{\frac{1}{2}}
\def\re{\mbox{Re}\, }
\def\im{\mbox{Im}\, }

\def\Mpl{M_{\rm Pl}}

\begin{document}
\vspace{-3em}
\rightline{ IFT-UAM/CSIC-26-67}
\vspace{2em}
\title{Gravity Decoupling and Axionic Shift Symmetries}
\author{Christian Aoufia} 
\email{christian.aoufia@estudiante.uam.es}
\author{Gonzalo F. Casas} 
\email{gonzalo.f.casas@csic.es}
\author{Fernando Marchesano} 
\email{fernando.marchesano@csic.es}
\affiliation{Instituto de F\'{\i}sica Te\'orica UAM-CSIC, c/ Nicol\'as Cabrera 13-15, 28049 Madrid, Spain}

\begin{abstract}

We analyse the role of approximate axionic shift symmetries in gravity-decoupling limits arising in type II Calabi--Yau compactifications. Associated with each shift symmetry there is an axionic string whose tension constrains the gradient of the K\"ahler potential, as expected in regimes where gravity becomes weakly coupled. The gradients of these tensions define vector fields on moduli space, analogous to those associated with BPS particle masses. Together, they characterise the evolution of different effective field theory sectors along asymptotic limits, encoding both their coupling to gravity and their kinetic mixing. By deriving upper bounds on the inner products of these vector fields, we show that they split into mutually orthogonal subsets, one of which decouples from gravity. Finally, we relate the Laplacian of certain axionic string tensions to a divergent moduli space curvature.

\end{abstract}
\maketitle
	
\section{Introduction}

In quantum gravity, infinite-distance limits in field space are expected to be accompanied by an infinite tower of asymptotically light states, as proposed by the Distance Conjecture \cite{Ooguri:2006in}. As the tower descends, the maximal cutoff of the infrared effective field theory (EFT) is lowered parametrically below the Planck scale, implying that gravity effectively decouples from part of the low-energy dynamics. From this perspective, gravity-decoupling limits are a natural consequence of exploring asymptotic regions in field space. Understanding how the various EFT sectors reorganise in these regimes, and how they remain coupled or decouple from one another, is therefore an important problem both from the viewpoint of effective field theory and the Swampland programme \cite{Vafa:2005ui}.

In Calabi--Yau string compactifications to four dimensions, these asymptotic regimes are typically accompanied by approximate axionic shift symmetries, degenerations of the moduli-space geometry, and light objects saturating the Bogomol'nyi-Prasad-Sommerfield (BPS) bound \cite{Bogomolny:1975de,Prasad:1975kr}. As stressed in \cite{Lanza:2020qmt,Lanza:2021udy,Lanza:2022zyg}, a key probe of these regimes is provided by BPS strings associated with the axionic shift symmetries. Their tensions encode geometric information about the compactification and are given by the first derivatives of the 4d EFT K\"ahler potential. In this sense, they fix the K\"ahler frame along the whole asymptotic region, as expected in regimes where gravity becomes weakly coupled \cite{Komargodski:2010rb,Adams:2011vw}.

At the same time, BPS particle masses define distinguished gradient vector fields on moduli space, dubbed charge vectors, which characterise the evolution of charged sectors along asymptotic trajectories, see e.g. \cite{Lee:2018spm,Calderon-Infante:2020dhm,Etheredge:2022opl,Etheredge:2023odp,Etheredge:2024tok}. Motivated by this parallel, in this work we study the vector fields defined as gradients of axionic string tensions (see also \cite{Grieco:2025bjy,Etheredge:2026rio}) and analyse their interplay with those associated with BPS particles. This provides a unified geometric framework to describe how different effective field theory sectors evolve near asymptotic limits, including both their coupling to gravity and their kinetic mixing.

Our analysis focuses on gravity-decoupling vector multiplet limits of 4d $\CN=2$ EFTs that arise from type II strings compactified on Calabi--Yau manifolds, following previous work \cite{Marchesano:2023thx,Marchesano:2024tod,Castellano:2024gwi,Castellano:2026bnx}. By deriving upper bounds on the inner products of gradient vector fields in moduli space associated with BPS strings and particles, we show that the asymptotic dynamics splits into mutually orthogonal vector-multiplet subsectors, as illustrated in figure \ref{fig:vectors}. Orthogonality implies a parametric suppression of both scalar and gauge kinetic mixing between these sectors. One of them, dubbed the core RFT, is the sector that decouples from the gravitational dynamics at the level of kinetic terms, while also accounting for divergences in the moduli-space curvature. Thus, the moduli-space geometry determines which sector decouples from gravity along the limit, geometrising how interactions between sectors are asymptotically suppressed.

We finally study the geometry of gravity decoupling through the Laplacians of string tensions. Using their relation to the Ricci operator on moduli space, we show that a divergent scalar curvature in asymptotic limits is bounded from above by the squared charge-to-tension ratio of a rigid axionic BPS string, paralleling analogous bounds for BPS particle charge-to-mass ratios \cite{Castellano:2024gwi}. Altogether, our results suggest that the physics of the rigid sectors emerging from gravity-decoupling limits is largely governed by the asymptotic geometry of field space. 

The rest of this letter is organised as follows. In the next section, we review 4d $\CN=2$ EFTs that arise from type II Calabi--Yau compactifications, as well as their BPS particle charge vectors. We next discuss how axionic shift symmetries describe charge vector fields for strings in moduli space, and their properties. We then relate the vector product to (gauge) kinetic mixing between vector multiplet sectors, and then derive a set of upper bounds that allow us to show the orthogonality relations depicted in figure \ref{fig:vectors}, both in the case of homogeneous K\"ahler potentials and in that of metric essential instantons. Finally, through the Laplacian of axionic string tensions, we relate the moduli scalar curvature to certain charge-to-tension ratios, and we draw our conclusions. 

\section{Type II Calabi--Yau compactifications and charge vectors}

Let us consider type II string theory compactified on a Calabi--Yau threefold (CY). The Lagrangian that describes the vector multiplet sector of the resulting 4d ${\cal N}=2$ EFT reads \cite{Ferrara:1988ff,Andrianopoli:1996cm,Lauria:2020rhc}
\begin{align}
\label{SVM}
S_{\rm 4d}^{\rm VM} & = \frac{1}{2\kappa_{4}^2} \int_{\mathbb{R}^{1,3}} R * \mathbbm{1} - 2 g_{i\bar{j}} dz^i \wedge * d\bar{z}^{\bar{j}} \\ & + \frac{1}{2} \int_{\mathbb{R}^{1,3}} \CI_{IJ} F^I \wedge * F^J + \CR_{IJ} F^I \wedge F^J ,
\nonumber
\end{align}
 where we follow similar conventions to \cite{Freedman_VanProeyen_2012}. Here, $z^i$ are the complex fields that parametrise the vector multiplet moduli space, and $g_{i\bar{j}}$ defines a special K\"ahler metric. For type IIB string theory compactified on a Calabi--Yau $Y$, this moduli space is given by its complex structure deformations and so $i = 1, \dots , h^{1,2}(Y) = n_V$. The K\"ahler potential describing the moduli space metric is globally defined in terms of its holomorphic three-form $\Omega$ as \cite{Strominger:1990pd}
\be
K = - \log \left(i\int_Y \Omega (z) \wedge \bar{\Omega} (\bar{z}) \right) = -\log \left( i \bar{\Pi}^A \eta_{AB} \Pi^B\right)
\label{Kpot}
\ee
where in the second equality we have chosen a basis of integer three-cycle classes $\Sigma_{A} \in H_3(Y,\mathbb{Z})$, whose Poincar\'e dual three-forms we denote by $\sig_A$, $A = 1, \dots , 2n_V +2$, and we have defined
\be
\Omega= \Pi^A \sig_A\, ,\qquad \eta_{AB} = - \int_Y \sig_A \wedge \sig_B\, .
\label{periodef}
\ee
Geometrically $\Pi^A$ is the integral of $\Omega$ over the homology class $- \eta^{AB}\Sigma_B$, where $\eta^{AB}$ is the inverse of $\eta_{AB}$, which is the topological intersection between $\Sigma_A$ and $\Sigma_B$. 

The 4d particles that source the field strengths $F^I$ and their magnetic duals $G_I \equiv \delta \CL / \delta F^I$, $I = 0, \dots , n_V$, arise from wrapping D3-branes on three-cycles of $Y$. If the three-cycle homology class is $\Sigma_\mathsf{q} = q^A \Sigma_A$, with $q^A \in \mathbb{Z}$, then the squared physical charge of this particle is
\be
\CQ^2_\mathsf{q} \equiv \oh \int_Y \sig_\mathsf{q} \wedge \star \sig_\mathsf{q} \equiv \oh q^A q^B \int_Y \sig_A \wedge \star \sig_B\, ,
\ee
where $\star$ stands for the Hodge star operator on $Y$. If the three-cycle is a special Lagrangian, then the particle is BPS and its mass is given by
\be
m_\mathsf{q} = |Z_\mathsf{q}|\, \Mpl\, , \ Z_\mathsf{q} \equiv e^{K/2} \langle \mathsf{q} , \mathsf{\Pi} \rangle \equiv e^{K/2} q^A \eta_{AB} \Pi^B\, , 
\label{massD3}
\ee
with $\Mpl = \sqrt{8\pi}/\kappa_4$, and where we have introduced the vector notation $\mathsf{\Pi} = (\Pi^1, \dots \Pi^{2n_V+2})^T$, similarly for $\mathsf{q}$. 

The physical charge matrix $\CQ^2_{AB} \equiv \oh \int_Y \sig_A \wedge \star \sig_B$ contains the same information as the gauge matrices $\CI_{IJ}$ and $\CR_{IJ}$. To find $\CQ^2_\mathsf{q}$, one needs to specify the Hodge star action on the basis elements $\sig_A$, which requires knowledge of the period vector $\mathsf{\Pi}$ and its derivatives. To see this, one may invoke the following identity \cite{Ceresole:1995ca,Palti:2017elp}
\be
\gamma_\mathsf{q}^2 \equiv \frac{\CQ^2_\mathsf{q}\Mpl^2}{m_\mathsf{q}^2} = 1 + 2 \lVert\vec \zeta_\mathsf{q} \rVert^2 .
\label{gamma}
\ee
Here $\gamma_\mathsf{q}$ stands for the charge-to-mass ratio of a BPS particle, and the {\it charge vector} $\vec{\zeta}_\mathsf{q}$ has entries
\be
\zeta_\mathsf{q}^a \equiv - g^{ab} \p_b \log (m_\mathsf{q}/\Mpl) ,
\label{zeta}
\ee
whose real index $a = 1, \dots, 2n_V$ is lowered with the real metric $g_{ab}$. These charge vectors have been used in the literature to classify asymptotic limits of infinite distance, see e.g. \cite{Lee:2018spm,Calderon-Infante:2020dhm,Etheredge:2022opl,Etheredge:2023odp,Etheredge:2024tok}. When applied to the present context, these works have focused on those charges $\mathsf{q}$ that host extremal BPS particles, since they realise the infinite tower of light states predicted by the Distance Conjecture \cite{Ooguri:2006in}. Due to their extremality, these particles have a charge-to-mass ratio $\gamma_\mathsf{q}$ of order one, and it turns out that this also extends to products of the form $\vec{\zeta}_\mathsf{p} \cdot \vec{\zeta}_\mathsf{q}$, with $\mathsf{p}$, $\mathsf{q}$ hosting extremal BPS particles \cite{Etheredge:2024tok}. 

This is, however, not so for BPS particles that do not saturate an extremality bound. Indeed, as stressed in \cite{Castellano:2024gwi}, certain BPS particles have a charge-to-mass ratio that blows up to infinity along such asymptotic limits. This has a clear meaning if one interprets the identity \eqref{gamma} as the no-force condition between two particles with the same charge $\mathsf{q}$, in which the attractive exchange of scalars and gravitons balances the repulsive force mediated by spin-1 fields \cite{Palti:2017elp}. When $\gamma_\mathsf{q} \gg 1$ the force of gravity is negligible with respect to the other two, and so the mutual interactions mediated by the U(1)$_\mathsf{q}$ sourced by $\mathsf{q}$ --- with field strength $\langle \mathsf{q} , \mathsf{F}\rangle$, $\mathsf{F} = (F^I,G_I)$ -- reproduce those of a rigid field theory \cite{Castellano:2026bnx}. Together with some additional decoupling conditions, this implies that this U(1) decouples from the graviphoton and the U(1)s sourced by extremal BPS particles \cite{Marchesano:2023thx,Marchesano:2024tod,Castellano:2024gwi,Castellano:2026bnx}. We then have a rigid field theory (RFT) that decouples from gravity. 

In the following, we argue that the necessary conditions for RFT gravity decoupling are not only reflected in the norms of the charge vectors, but also in the angles between them. A key ingredient to attain this picture will be the presence of axionic shift symmetries that become exact at the limit endpoint. As a consequence, there are a number of BPS axionic strings in the spectrum, each with its own charge vector. The properties of these vectors imply that the different field theory sectors that appear in gravity decoupling limits become orthogonal to each other when mapped to the tangent field space, providing a geometric EFT picture of gravity decoupling.

\section{Strings and axionic shift symmetries}

A well-known framework to describe the physics of gravity-decoupling limits is special geometry, see e.g. \cite{Andrianopoli:1996cm}. In particular, with a choice of special coordinates, all the couplings in \eqref{SVM} are specified in terms of a prepotential. Alternatively, one may exploit the Nilpotent Orbit Theorem of Schmid \cite{Schmid:1973cdo} to locally describe the period vector $\mathsf{\Pi}$ around certain moduli space singularities:
\be
    \mathsf{\Pi} = 
   e^{T^i P_i} \left( {\bf a}_0 + \sum_{r_n} {\bf a}_{r_1\dots r_{N}}\, e^{2\pi i r_i T^i} \right)\, ,
\label{periodv}
\ee
where $T^i = a^i + i t^i$ describe the coordinates transverse to the singularity, located at $t^i \to \infty$ and surrounded as $a^i \to a^i +1$. Such periodicity implements a monodromy transformation $\mathsf{\Pi} \to e^{P_i} \mathsf{\Pi}$ on the period vector, where $e^{P_i}$ are commuting integer unipotent matrices. Finally, $r_i \in \mathbb{N}$ and ${\bf a}_{0}, {\bf a}_{r_1\dots r_{N}} \in \mathbb{C}^{2n_V+2}$.\footnote{Singularities have complex codimension $N\leq n_V$, parametrised by the transverse fields $T^i$, $i=1,\dots, N$, while ${\bf a}_{0}, {\bf a}_{r_1\dots r_{N}}$ are holomorphic functions of the parallel fields. Our discussion assumes no significant kinetic mixing between both sets of fields.} This and related results apply to a large class of asymptotic limits in Calabi--Yau moduli spaces, and have been used in \cite{Grimm:2018ohb,Grimm:2018cpv,Corvilain:2018lgw} (and more recently in \cite{Hassfeld:2025uoy,Monnee:2025ynn}) to understand and classify the physics of the Distance Conjecture along 4d $\CN=2$ vector multiplet infinite distance limits, in type II Calabi--Yau settings. In particular, the type of singularity is defined via the index $d=0,1,2,3$ such that $P^{d}{\bf a}_0 \neq 0$ but $P^{d+1}{\bf a}_0 =0$, where $P=c^i P_i$, for some choice of $c^i \in \mathbb{R}_+$. 

For the following discussion, the most relevant fact about \eqref{periodv} is that the approximate period and the corresponding K\"ahler potential
\be
\mathsf{\Pi}_{\rm pol} \equiv e^{T^i P_i} {\bf a}_0 , \qquad K_{\rm pol} = -\log \left( i \bar{\Pi}_{\rm pol}^A \eta_{AB} \Pi^B_{\rm pol}\right) ,
\label{Pipol}
\ee 
display a very specific structure. Due to the nilpotency of the $P_i$ one has that $\mathsf{\Pi}_{\rm pol}$ is a polynomial in the $T^i$ of degree $d$, and due to the fact that $a^i \to a^i +1$ is a discrete symmetry, we have that $P_i^T \eta + \eta P_i =0$. As a result $K_{\rm pol}$ displays the following axionic shift symmetries
\be
a^i \to a^i + \lambda^i, \qquad \lambda^i \in \mathbb{R},
\ee
in those regimes where $t^i \gg e^{-2\pi t^j} \forall i,j$. In the following, we will restrict to those regions, since they justify the expansion in \eqref{periodv}. More precisely, we will have in mind the so-called growth sector limits
\be
t^i = t^i_0+ e^i_0 \rho + e^i_1 \rho^{\a_1} + e^i_2 \rho^{\a_2} + \dots 
\label{growth}
\ee
where $\rho \to \infty$ and ${\bm e}_m$, $m =0 ,1,2, \dots$ are vectors with $n_V$ non-negative integer $e^i_m$ entries, such that $\sum_i e^i_m \cdot e^i_n = 0$, $\forall m \neq n$, and $1> \a_1 > \ldots> \a_m > \a_{m+1}> \dots$

These approximate shift symmetries imply the presence of a set of axionic BPS strings in the 4d EFT spectrum, whose tension can be approximated as \cite{Lanza:2020qmt,Lanza:2021udy,Lanza:2022zyg}
\be
\CT_{\bm e} \equiv e^i \CT_i = -\oh e^i \p_{t^i} K \, \Mpl^2,
\label{Tstring}
\ee
where the vector ${\bm e}$ with $n_V$ integer-valued entries represents the string charge. If all the entries of this vector are non-negative, the exponential corrections to $\mathsf{\Pi}_{\rm pol}$ will decrease as we approach the string core along its 4d solution. More precisely, in terms of the transverse 4d coordinate $z_{4d} = \exp(2\pi (-\rho + i \sigma))$ one finds the following BPS solution around a string located at $z_{4d} =0$
\be
t^i = t^i_0 + e^i \rho, \quad a^i = \sigma e^i ,
\label{flow}
\ee
which is a particular case of the saxionic trajectory \eqref{growth}. Axionic BPS strings satisfying this property were dubbed EFT strings in \cite{Lanza:2020qmt,Lanza:2021udy,Lanza:2022zyg}, and shown to be extremal objects whose physics is captured by $K_{\rm pol}$. Charges ${\bm e}$ with some negative entry describe non-EFT strings, for which the solution \eqref{flow} is not reliable near their core. This happens, for instance, in asymptotic regions where $K_{\rm pol}$ yields a degenerate metric along some field direction $T^i \propto e^i$, dubbed limits with metric essential instantons \cite{Bastian:2021eom,Bastian:2021hpc}. In that case, and depending on the exponential terms, one may or may not have an approximate axionic shift symmetry for the $K_{\rm pol}$-degenerate directions. Whenever one does, one may compute the BPS axionic string tension as in \eqref{Tstring}, and then consider the charge vector 
\be
\xi_{\bm e}^a \equiv - g^{ab} \p_b \log ( \CT_{\bm e}/\Mpl^2) .
\label{xi}
\ee
In the following, we will work in Planck units. Due to the no-force condition for BPS strings, one has $\lVert\vec \xi_{\bm e} \rVert = \gamma_{\bm e}$ \cite{Lanza:2020qmt}, namely the string charge-to-tension ratio. Then, just like for particles, along a given limit one will have BPS extremal and non-extremal strings. The former have $\gamma_{\bm e} \sim \CO(1)$ and include EFT strings, for which their products also satisfy $\vec{\xi}_{\bm e} \cdot \vec{\xi}_{\bm f} \sim \CO(1)$ \cite{Grieco:2025bjy}. The latter have $\gamma_{\bm e} \gg \CO(1)$ which is equivalent to require
\be
|\vec{e} \cdot \vec{\nabla} K| \ll 2\lVert \vec{e} \rVert , \quad \text{with} \quad \vec{e} \equiv ({\bm e}, {\bm 0})^T.
\label{RFTmetric}
\ee
This is nothing but the RFT condition defined in \cite{Castellano:2024gwi,Castellano:2026bnx}, along the saxionic field direction $\vec{e}$. In contrast to that reference, where it is related to the ratio of a BPS magnetic state, here we relate it to a string ratio. Both conditions coincide in decompactification limits to 5d \cite{Blanco:2025qom}.

The period vector \eqref{periodv} not only implies a simple description for axionic BPS string tensions, but also for the mass of the leading tower of extremal states. Indeed, in this framework, the states of such a tower satisfy $P_i \mathsf{q} = 0, \forall i$ \cite{Hassfeld:2025uoy,Monnee:2025ynn}, and as a result for its lightest state $\mathsf{q}_*$ we have
\be
m_* = e^{K/2} a_* \Mpl \implies \vec{\zeta}_* = - \oh \vec\nabla K ,
\label{mstar}
\ee
where $a_* \equiv |\langle \mathsf{q}_*, {\bf a}_0\rangle|$ is a positive real constant. One can also check that the norm of this charge vector is finite
\be
\lVert \vec{\zeta}_* \rVert \simeq \sqrt{d/2}
\label{normstar}
\ee
where $d>0$ is defined above \eqref{Pipol} and coincides with the degree of $\exp(-K_{\rm pol})$. This $\CO(1)$ norm reflects the fact that the BPS tower is also extremal, as previously noted.

Notice that the above expressions imply a specific choice of K\"ahler frame, since they connect $\vec{\nabla}K$ to physical quantities like BPS string tensions. One can trace back this lack of invariance under general K\"ahler transformations $K \to K + F + \bar{F}$ with $F$ holomorphic to the particular form of the period vector \eqref{periodv}, which is not invariant under $\Omega \to e^{-F} \Omega$ unless $F$ is a constant. That this choice of K\"ahler potential can be made for the whole asymptotic region may be surprising at first, but it precisely matches the general requirement that the K\"ahler form of field space must be exact in order to couple supergravity to a rigid theory \cite{Komargodski:2010rb,Adams:2011vw}. In the present setting, this exactness can be linked to the presence of explicit axionic shift symmetries in the K\"ahler potential. 

\section{Vector products and mixing}

An interesting observation is that, in the presence of axionic shift symmetries, charge vectors provide a geometric description of the kinetic mixing between different sectors. Indeed, the charge vector of a BPS axionic string with tension $\CT_{\bm e}$ reads
\be
\vec{\nabla} {\CT}_{\bm e} = -\vec{e}, \qquad \vec{\xi}_{\bm e} = \frac{\vec{e}}{{\CT}_{\bm e}} \, .
\label{gradT}
\ee 
This not only yields $\lVert\vec \xi_{\bm e} \rVert = \lVert\vec{e} \rVert /\CT_{\bm e} = \gamma_{\bm e}$, but also
\be
\cos \theta_{(\vec{e}, \vec{f})} \equiv \frac{\vec{e} \cdot \vec{f}}{\lVert\vec{e} \rVert \lVert\vec{f} \rVert} = \frac{\vec{\xi}_{\bm e} \cdot \vec{\xi}_{\bm f}}{\lVert\vec{\xi}_{\bm e} \rVert \lVert\vec{\xi}_{\bm f} \rVert}.
\label{anglexi}
\ee
That is, the vector products between different string charge vectors determine the kinetic mixing between field directions, and more precisely, their relative angles. Again, this is not only true for EFT strings, but for any pair of BPS axionic strings. 

In the case of BPS particles, their products determine the gauge kinetic mixing. Indeed, one may show that
\be
|Z_\mathsf{p} Z_\mathsf{q}| + 2\vec{\nabla} |Z_\mathsf{p}| \cdot \vec{\nabla} |Z_\mathsf{q}| = \cos \varphi_\mathsf{pq} \CQ^2_{\mathsf{p}\mathsf{q}} + \oh \sin\varphi_\mathsf{pq} \langle \mathsf{p},\mathsf{q} \rangle,
\label{mixQ1}
\ee
where $\CQ^2_{\mathsf{p}\mathsf{q}} \equiv \CQ^2_{AB} p^A q^B$ represents the gauge kinetic mixing between U(1)$_\mathsf{p}$ and U(1)$_\mathsf{q}$, and $\exp( i \varphi_\mathsf{pq}) = \arg Z_\mathsf{p}\bar{Z}_\mathsf{q}$. As a result, one finds the following relation
\be
\label{mixQ2}
 \frac{1+ 2 \vec{\zeta}_\mathsf{p} \cdot \vec{\zeta}_\mathsf{q}}{(1+ 2\lVert\vec{\zeta}_\mathsf{p}\rVert^2)^\oh (1+ 2\lVert\vec{\zeta}_\mathsf{q}\rVert^2)^\oh} = \frac{{\rm Re} \left[e^{-i \varphi_\mathsf{pq}}( \CQ^2_{\mathsf{p}\mathsf{q}} + \frac{i}{2} \langle \mathsf{p},\mathsf{q} \rangle) \right]}{\CQ_{\mathsf{p}} \CQ_{\mathsf{q}}}.
\ee

Finally, one may consider products between string and particle charge vectors. If ${\bm e}$ corresponds to an EFT string, these compute the exponential BPS mass variation along \eqref{flow}, namely
\be
\lambda_{{\bm e},\mathsf{q} } \equiv \vec{e} \cdot \vec{\zeta}_\mathsf{q}/\lVert\vec{e} \rVert = \vec{\xi}_{\bm e} \cdot \vec{\zeta}_\mathsf{q}/ \lVert\vec{\xi}_{\bm e} \rVert,
\label{mixQmix}
\ee
 which upon normalisation of $\vec{\zeta}_\mathsf{q}$ gives
\be
 \cos \phi_{({\bm e},\mathsf{q})} \equiv \frac{\vec{\xi}_{\bm e} \cdot \vec{\zeta}_\mathsf{q}}{\lVert\vec{\xi}_{\bm e} \rVert \lVert\vec{\zeta}_\mathsf{q}\rVert} .
 \label{mixQmix2}
\ee
Similarly, the product $\vec{\xi}_{\bm e} \cdot \vec{\xi}_{\bm f}$ can be interpreted in terms of the exponential variation of $\CT_{\bm f}$ along \eqref{flow}. 

To sum up, one finds that the kinetic mixing, gauge kinetic mixing and mass variations along string flows can all be captured in terms of charge vector products. Because the norms of these vectors are related to the charge-to-tension ratio of BPS strings and particles, this already suggests suppressed mixing terms between extremal and certain rigid sectors. To see this more clearly, it proves useful to establish certain bounds on the scalar products of charge vectors, as we discuss next.

\section{Bounds on vector products}

Given the above definitions, it is easy to see that 
\be
\vec{\xi}_{\bm e} \cdot \vec{\zeta}_* = 1,
\label{stringstar}
\ee
for any axionic string charge ${\bm e}$. Interestingly, for infinite distance limits with leading direction ${\bm e}$, one can identify the species scale $\Lambda_{\rm sp}$ with $\sqrt{\CT_{\bm e}}$ \cite{Marchesano:2022axe}. From there, one precisely recovers the pattern observed in \cite{Castellano:2023stg,Castellano:2023jjt} relating the leading tower to the species scale, for this particular setting. More generally, the fact that in this context the relation extends to all axionic BPS strings, both extremal and non-extremal, reveals a simple geometric distinction between the two kinds. Indeed, it follows from \eqref{stringstar} and \eqref{normstar} that the angle between $\vec{e}$ and $\vec{\nabla} K$ reads
\be
\cos \phi_{({\bm e},*)} \simeq \sqrt{\frac{2}{d}}\frac{1}{ \gamma_{\bm e}} .
\label{anglestar}
\ee
As a result for extremal strings $\cos \phi_{({\bm e},*)} \simeq \CO(1)$, while for rigid ones $\cos \phi_{({\bm e},*)} \to 0$ as we proceed along \eqref{growth}. In other words, the rigid field directions represented by $\vec{e}$ become orthogonal to $\vec{\nabla} K$, in the specific K\"ahler frame that we are working in. Excursions along these rigid field directions represent variations of $K$ that are negligible compared to those generated by EFT string flows. This fits well with the standard picture in which, whenever we have a rigid limit we can express the K\"ahler potential as 
\be
K = - \log \left( \CK_0 - \CK_{\rm rigid}\right)\, ,
\label{Kexp}
\ee
where $\CK_0$ is independent of the rigid fields and $\CK_0 \gg \CK_{\rm rigid}$. This hierarchy justifies the expansion
\be
K = - \log \CK_0 + \frac{\CK_{\rm rigid}}{\CK_0} + \dots
\label{Kexp2}
\ee
from where we can identify $K_{\rm rigid}$ with the second term. It follows that the excursions of those fields contained in $\CK_0$ may translate into a variation of the K\"ahler potential of Planckian size, while for rigid fields the variation is of order $m_* \simeq \Mpl/\CK_0^{1/2}$. In the presence of shift symmetries, this can be translated to relative angles between charge vectors. Notice in particular that the suppression increases for larger values of $\gamma_{\bm e}$. 

More generally, one can draw a geometric picture describing extremal and rigid sectors by looking at the relative angles between charge vectors. Let us first focus on those that correspond to axionic strings. Assuming the presence of axionic shift symmetries along the saxionic directions $\vec{e} = (\bm{e}, 0)$ and $\vec{f} = (\bm{f}, 0)$, one finds that their product reads 
\begin{equation}
   \vec{e}\cdot \vec{f} = 2{\CT}_{\bm e} {\CT}_{\bm f} + (\vec{e} \cdot \vec{f})_{\rm rig}\ ,
   \label{metric}
\end{equation}
where the rhs product is computed with the rigid metric 
\be
g_{ij}^{\rm rigid} \equiv - \oh e^K \p_{t^i} \p_{t^j} e^{-K} .
\label{rmetric}
\ee
As a result, the product of two string charge vectors is
\begin{equation}
   \vec \xi _{\bm e}\cdot \vec \xi_{\bm f} = \left( 2+ \frac{(\vec{e}\cdot \vec{f})_{\rm rig}}{{\CT}_{\bm e} {\CT}_{\bm f}}\right) .
   \label{prodstst}
\end{equation}

When both vectors represent EFT strings, the rigid contribution is either comparable or subleading with respect to the first one. Hence
\be
 \vec{\xi}_{\bm e} \cdot \vec{\xi}_{\bm f} \lesssim \CO(1) .
\label{boundstst}
\ee
For rigid strings, orthogonality with respect to $\vec{\nabla}K$ implies that the rigid contribution is dominant:
\be
\vec{\xi}_{\bm e} \cdot \vec{\xi}_{\bm f} \simeq +\frac{(\vec{e} \cdot \vec{f})_{\rm rig}}{{\CT}_{\bm e} {\CT}_{\bm f}} , \quad \cos \theta_{(\vec{e}, \vec{f})} \simeq + \frac{(\vec{e} \cdot \vec{f})_{\rm rig}}{\lVert\vec{e} \rVert_{\rm rig}\lVert\vec{f} \rVert_{\rm rig}} .
\label{boundststring}
\ee
From here one can deduce the following bound 
\be\label{eq:stringbound}
\vec{\xi}_{\bm e} \cdot \vec{\xi}_{\bm f} \lesssim \min (\CT_{\bm e}^{-1}, \CT_{\bm f}^{-1}).
\ee
 This implies that, in the rigid sector, the products $\vec{\xi}_{\bm e} \cdot \vec{\xi}_{\bm f}$ can diverge asymptotically, as for instance happens with their charge-to-tension ratios $\gamma_{\bm e} = \lVert\vec{\xi}_{\bm e} \rVert$. More precisely, for rigid strings whose tension can be computed as ${\CT}_{\bm f} = - \oh \vec{f} \cdot \vec{\nabla} K_{\rm pol}$, as happens in large complex structure (LCS) limits, the norm $\lVert\vec{\xi}_{\bm f} \rVert$ diverges polynomially in the saxions $t^i$. For those that arise from a metric instanton sector, $\lVert\vec{\xi}_{\bm f} \rVert$ will instead grow exponentially with the $t^i$, see the next sections for more details. 

The bound \eqref{eq:stringbound} also applies when $\bm{e}$ corresponds to an EFT string charge and $\bm{f}$ to a rigid one. Then, because in practice the rigid charge is the lightest, we can replace the rhs by $\CT_{\bm e}^{-1}$. The angle between both charge vectors is then bounded as
\be
\cos \theta_{(\vec{e}, \vec{f})} \lesssim (\CT_{\bm e} \gamma_{\bm f})^{-1}. 
\label{anglestmix}
\ee
If $\bm{f}$ corresponds to a metric instanton sector, then $\gamma_{\bm f}$ will grow exponentially faster than $\CT_{\bm e}^{-1}$, and so $\cos \theta_{(\vec{e}, \vec{f})} \to 0$. If, on the other hand, $\gamma_{\bm f}$ grows polynomially along the limit, the charge vectors will become orthogonal only for selected extremal-rigid pairs $(\vec{\xi}_{\bm e}, \vec{\xi}_{\bm f})$. In particular, the bound \eqref{anglestmix} favours orthogonality for the rigid subsector whose charge-to-tension ratio $\gamma_{\bm f}$ is most divergent, which corresponds to the RFT subsector that can decouple from gravity, also known as \emph{core RFT} \cite{Castellano:2024gwi,Blanco:2025qom}. 

Crucially, for selected EFT string directions \eqref{boundstst} still applies. Indeed, as noted above, one may interpret \eqref{boundstst} as that, along the EFT string flow \eqref{flow}, all EFT string tensions evolve as some power of $\CT_{\bm e} \simeq 1/\rho$. If a rigid string tension $\CT_{\bm f}$ is computed from $K_{\rm pol}$ then it also depends polynomially on $\CT_{\bm e}$, and so along the EFT string flow generated by ${\bm e}$ we have $\vec{\xi}_{\bm e}\cdot \vec{\xi}_{\bm f} \lesssim \CO(1)$. The same bound applies to those EFT strings with charges $\bm{e}_\a$ such that $\CT_{\bm{e}_\a} \simeq \CO(1)$ along this same flow. In both cases $\cos \theta_{(\vec{e}_{\a}, \vec{f})} \lesssim 1/\gamma_{\bm{f}} \to 0$ for any rigid string charge $\bm{f}$. In other words, all rigid string charge vectors become asymptotically orthogonal to the field directions $\{\vec{e}, \vec{e}_\a\}$, which we denote as the \emph{SDC sector}.\footnote{The reason for this name is that, as can be deduced from Appendix B of \cite{Marchesano:2023thx} and the results of \cite{Marchesano:2024tod}, the saxions $(t,t^\a)$ that correspond to $(\bm{e},\bm{e}_\a)$ control the scaling of the SDC towers.}

Let us now consider mixed vector products $\vec{\xi}_{\bm e} \cdot \vec{\zeta}_\mathsf{q}$ between string and particle charge vectors. Just like for strings, one may consider two kinds of particle charge vectors. The first corresponds to a central charge of the form $Z_\mathsf{q} \simeq e^{K/2} \langle \mathsf{q} , \mathsf{\Pi}_{\rm pol} \rangle$, and the second to exponentially-suppressed periods that arise from a metric essential instanton sector \cite{Castellano:2026bnx}. In the first case let us assume that 
\be
m_{\mathsf{p}} \simeq e^{K/2} |T_d^n| \Mpl,
\ee
where $T_d \equiv d_i T^i$. Then one finds
\be
\vec{\xi}_{\bm e} \cdot \vec{\zeta}_{\mathsf{p}} \simeq 1 + n d_ie^i\frac{t_d}{{\CT}_{\bm e} |T_d|^2} .
\ee
The rhs is $\CO(1)$ when ${\bm e}$ corresponds to the leading direction of the limit. For other EFT string charges we instead have the bound
\be
\vec{\xi}_{\bm e} \cdot \vec{\zeta}_{\mathsf{p}} \lesssim {\CT}_{\bm e}^{-1} ,
\label{generalbound}
\ee
which also holds for exponentially-suppressed masses. So we find a similar pattern to that obtained for string charge vectors. All rigid U(1)$_\mathsf{p}$'s become orthogonal to $\vec{\xi}_{\bm e}$ when $\bm{e}$ corresponds to the leading direction of the limit or $\CT_{\bm{e}} \simeq \CO(1)$. For other EFT string charges, the orthogonality depends on the degree of divergence of $\gamma_\mathsf{p}$.
 
To sum up, we find that charge vectors $\vec\xi_{\bm f}$ and $\vec{\zeta}_\mathsf{q}$, that respectively represent rigid field directions and rigid U(1)s, become asymptotically orthogonal to the SDC sector. Orthogonality with respect to other EFT string charge vectors depends on the degree of divergence of $\lVert\vec{\xi}_{\bm f} \rVert$ and $\lVert\vec{\zeta}_\mathsf{q} \rVert$, and points towards those rigid sectors that decouple from gravity. As stressed in \cite{Castellano:2024gwi,Castellano:2026bnx}, a necessary condition to decouple a vector multiplet from gravity is that both its electrically and magnetically charged states have diverging charge-to-mass ratios. To see how this translates into the geometry of charge vectors, let us consider a magnetic state $\mathsf{q}$ and its electric dual state $\mathsf{p}$ in the sense of $\langle \mathsf{p}, \mathsf{q}\rangle \neq 0$, and such that $\exp( i \varphi_{\mathsf{p}\mathsf{q}}) = i$. Applying this to \eqref{mixQ1} while assuming that $\gamma_{\mathsf{p}}, \gamma_{\mathsf{q}} \gg 1$, one arrives at
\be
2\CQ_{\mathsf{p}} \CQ_{\mathsf{q}} = \langle \mathsf{p},\mathsf{q} \rangle \iff \vec{\zeta}_\mathsf{p} \parallel \vec{\zeta}_\mathsf{q}. 
\label{pqond}
\ee
That is, to recover the usual field theory relation between dual electro-magnetic couplings $g_e g_m \in \mathbb{Z}$ one needs parallel charge vectors. In fact, in order to decouple a vector multiplet associated with a rigid field direction $T^i \propto f^i$ one should require a structure of the form $\vec{\xi}_{\bm f} \parallel \vec{\zeta}_{\mathsf{p}_{\bm f}} \parallel \vec{\zeta}_{\mathsf{q}_{\bm f}}$. This is stronger than requiring each of these vectors to be orthogonal to $\vec\zeta_*$ or the SDC sector, and it is closer to the condition of orthogonality with respect to all charge vectors for EFT strings and extremal particles. To show how this more detailed structure arises, it proves useful to consider separately the two main setups that arise in asymptotic limits.

\section{The homogeneous case}

Let us consider the piece of the K\"ahler potential that arises at the polynomial level
\be
K_{\rm pol} = - \log ({\cal G}), 
\label{Kpol}
\ee
where $\CG\equiv \CG(t^i)$ is a polynomial function in the saxions $\{t^i\}$. In the LCS regime, one can approximate it by a homogeneous polynomial of degree $d=3$. Then one has
\be
\vec{\zeta}_* = - \oh \vec\nabla K_{\rm pol} \simeq ({\bm t}, {\bm 0})^T, 
\label{LCSstar}
\ee
where the components of ${\bm t}$ are the saxion coordinates $t^i$. This in turn implies that 
\be
\vec{\zeta}_* \cdot \vec{\nabla} \log f_{Q} \simeq + t^i \p_{t^i} \log f_{Q} \simeq  + Q.
\label{Qbound}
\ee
where $f_{Q}$ is a degree $Q$ function on the $t^i$. This holds for all the tensions and masses of BPS strings and particles in the LCS regime. 

The bound on the product \eqref{Qbound} means that all BPS strings and particles with diverging ratio $\gamma$ will become orthogonal to $\vec{\nabla} K_{\rm pol}$ along an asymptotic limit with rigid sectors, just as they become orthogonal to the leading direction $\vec{\xi}_{{\bm e}_0} \propto {\bm e}_0$ of the limit \eqref{growth}, as previously discussed.

To see how they behave with respect to other extremal directions, let us consider the rigid metric 
\be
g_{ij}^{\rm rig} = - \CG^{-1} \oh \p_{t^i} \p_{t^j} \CG .
\label{grigLCS}
\ee
Following \cite{Marchesano:2023thx,Marchesano:2024tod,Blanco:2025qom} we define the core RFT sector as those field directions ${\bm f}$ that are exactly orthogonal to the leading direction ${\bm e}_0$ with respect to \eqref{grigLCS}, while the whole set of rigid directions is dubbed extended RFT. As mentioned, the core RFT directions display those BPS strings with the largest tension-to-mass ratio. For instance, in the particular case of EFT string limits \eqref{flow} one has that $\gamma_{\bm f}^2 \simeq \CG \simeq \CT_{\bm{e}_0}^{-w}$ for a string that belongs to the core RFT, where $w=1,2,3$ is the scaling weight defined in \cite{Lanza:2021udy}. For extended RFT vectors outside of the core RFT, we instead have that $\gamma_{\bm f}^2 \simeq \CG\CT_{\bm{e}_0} \simeq \CT_{\bm{e}_0}^{-w+1}$, for $w=2,3$.

Let us now use \eqref{anglestmix} to see the interplay between these two RFT sectors and the sector of EFT strings, which contains the SDC subsector defined above. For simplicity, let us focus on EFT string LCS limits, whose spectrum of EFT strings was analysed in Appendix B of \cite{Marchesano:2023thx} in terms of the scaling weight $w$. For $w=3$ the SDC sector reduces to the direction of the limit $\bm{e}_0$, and there are $n_V-1$ independent EFT string charges ${\bm e}_i$ such that $\CT_{{\bm e}_i} \simeq \CT_{{\bm e}_0}$. For a generic extended RFT vector we have $\cos \theta_{(\vec{e}_i, \vec{f})} \simeq \CO(1)$, while if $\bm{f}$ belongs to the core RFT we have $\cos \theta_{(\vec{e}_i, \vec{f})} \to 0$ instead. For $w=1$ limits the only string tension that tends to zero is $\CT_{{\bm e}_0}$. Hence the SDC and EFT string sectors coincide, just like the core and extended RFT sectors do, and so $\cos \theta_{(\vec{e}_i, \vec{f})} \to 0$ for any rigid direction $\bm{f}$. Finally, $w=2$ limits occur in elliptically fibred CYs. Their SDC sector comprises the direction of the limit $\bm{e}_0$ and the field direction $\bm{e}_E$ that controls the generic elliptic fibre. The core RFT becomes orthogonal to both of them but not to other EFT string charges $\bm{e}_i$ with $\CT_{{\bm e}_i} \simeq \CT_{{\bm e}_0}$, which correspond to vertical divisors. Still, one can split the set of charge vectors into a sum of orthogonal subsectors: the SDC sector, the core RFT and the rest of the extended RFT.

This last result relies on the orthogonality structure that also occurs within the extended RFT. Indeed, one can see that for EFT string limits $(\vec{e} \cdot \vec{f})_{\rm rig} \sim (\CG \CT_{\bm{e}_0})^{-1}$ for two generic vectors in the extended RFT, while it goes like $\CG^{-1}$ if at least one of them belongs to the core RFT. Plugging this behaviour into \eqref{boundststring}, one can see that core RFT directions become asymptotically orthogonal to the remaining extended RFT directions, suppressing the kinetic mixing between both sectors. This lesson extends to the more general set of limits \eqref{growth}, where the asymptotic dependence of these vector products is more involved. Ultimately, one can trace back the orthogonality between core and extended RFT directions to the fact that one is generating a hierarchy of eigenvalues along these limits:
\be
g_{ij}^{\rm core RFT} \ll g_{ij}^{\rm ext RFT},
\ee
while the contracted vectors $\vec{e}$, $\vec{f}$ have integer entries. 

Turning to particle charge vectors, one may follow \cite{Castellano:2024gwi} and consider certain magnetic states with a choice of charge $\mathsf{q}_{\bm f}$ such that $m_{\mathsf{q}_{\bm f}} \simeq \CT_{\bm f}/m_*$, where we have used that those field directions $T^i$ whose metric is dominated by $K_{\rm pol}$ can be identified with special coordinates. Translated into charge vectors one then finds
\be
 \vec{\zeta}_{\mathsf{q}_{\bm f}}|_{\rm sax} \simeq \vec{\xi}_{\bm f} - \vec{\zeta}_* ,
 \label{magvec}
\ee
where we have neglected subleading axion-dependent terms and on the lhs we have projected $\vec{\zeta}_{\mathsf{q}_{\bm f}}$ into its saxionic entries. Since those are the only ones that contribute to products with string charge vectors, we have
\begin{equation}
   \vec \xi _{\bm e}\cdot \vec{\zeta}_{\mathsf{q}_{\bm f}} \simeq 1- 2 \frac{(\vec{e}\cdot \vec{f})_{\rm rig}}{\CT_{\bm e}\CT_{\bm f}} ,
\end{equation}
and so the discussion below \eqref{prodstst} extends to this case. If ${\bm e}$ belongs to the SDC sector or represents a generic extended RFT direction and $\mathsf{q}_{\bm f}$ a core RFT one, then $\vec \xi _{\bm e}$ and $\vec \zeta_{\mathsf{q}_{\bm f}}$ will become orthogonal asymptotically. More generally, at the polynomial level one has that the axionic components of $\vec \zeta_{\mathsf{q}_{\bm f}}$ are negligible compared to $ \vec{\zeta}_{\mathsf{q}_{\bm f}}|_{\rm sax}$, while due to \eqref{Qbound} one has that $\lVert\vec{\zeta}_* \rVert \ll \lVert \vec{\zeta}_{\mathsf{q}_{\bm f}}\rVert_{\rm sax}$ for rigid BPS particles. As a result, the two vectors $\vec{\zeta}_{\mathsf{q}_{\bm f}}$ and $\vec{\xi}_{\bm f}$ will asymptotically align for rigid sectors. 

Hence, together with the condition \eqref{pqond}, one arrives at the structure $\vec{\xi}_{\bm f} \parallel \vec{\zeta}_{\mathsf{p}_{\bm f}} \parallel \vec{\zeta}_{\mathsf{q}_{\bm f}}$. This typically holds asymptotically for core RFT sectors, generalising the alignment of charge vectors observed in \cite{Reece:2025zva} to 4d $\CN=2$ EFTs with axionic shift symmetries. 

\section{Metric essential instantons}
Limits with metric essential instantons have a K\"ahler potential with the following structure \cite{Bastian:2021hpc}
\be
K = - \log \left( \CK_{\rm pol} - \CK_{\rm inst}\right)\, ,
\label{Kmetric}
\ee
where $\CK_{\rm pol}$ is a polynomial function of degree $d$ in the saxions $t^k$, and $\CK_{\rm inst}$ contains the exponential corrections. The piece $K_{\rm pol} = -\log \CK_{\rm pol}$ leads to a degenerate metric, and so at least some terms within $\CK_{\rm inst}$ must be considered in order to lift such a degeneracy. Following \cite{Castellano:2026bnx}, let us assume that such terms are of the form
\be
 \CK_{\rm inst} = Q_{\mu\nu} z^\mu \bar{z}^\nu + \im (B_\mu z^\mu) + \dots 
 \label{Kess}
\ee
where $z^\mu = \exp(2\pi i T^\mu)$ depend on the field directions that do not appear in $\CK_{\rm pol}$, and $Q_{\mu\nu}, B_\mu$ are polynomial functions of the $t^k$. An interesting observation is that, for infinite-distance limits, these polynomials are always of the same or lower degree than $\CK_{\rm pol}$, such that $\CK_{\rm pol} \gg \CK_{\rm inst}$ in the moduli space region around $t^k = \infty$ \cite{Bastian:2021eom}.\footnote{We thank D. van de Heisteeg for discussions on this point.} 

Let us for simplicity assume that $\CK_{\rm pol}$ depends on a single saxionic field $t^1 = \im T^1$, while there is a single relevant exponential in \eqref{Kess}: $z^\mu = \exp (2\pi i (r_1^\mu T^1 + r_2^\mu T^2))$, $r_i^\mu \in \mathbb{N}$. That is, the K\"ahler potential is
\be
K = - \log \left((t^1)^d - Q |z^\mu|^2 - \im(B z^\mu) +\dots \right),
\ee
with $Q$ and $B$ in general polynomial functions of $(t^1, t^2)$. The metric in the basis $dT^i$ at leading order reads $ g_{T^1\bar{T}^1}\simeq d/(4(t^1)^2)$ and
\be
g_{T^i\bar{T}^2} = -\frac{\partial_{T^i}\partial_{\bar{T}^2}\left({\rm Im}(Bz^{\mu})\right)}{(t^1)^d} + \mathcal{O}(|z^{\mu}|^2)\, ,
\ee
from where one can see that the axionic shift symmetry for $\re T^2$ is broken unless $B=0$. In that case, the leading order takes the shape
\begin{equation}
    g_{T^i\bar{T}^2} = -\frac{\partial_{T^i}\partial_{\bar{T}^2}\left(Q |z^{\mu}|^2\right)}{(t^1)^d} + \mathcal{O}((t^1)^{-d-1})\, ,
\end{equation}
and the string tensions read
\be
\begin{split}
&\CT_{\bm 1} = \frac{d}{2 t^1} + \CO(z^{\mu}) , \\ &\CT_{\bm 2} =- \frac{1}{2(t^1)^d} \partial_{t^2}(Q |z^\mu|^2) + \CO(|z^\mu|^2) . 
\end{split}
\ee 
Their charge vectors take the form
\be
\vec \xi_{\bm 1}\simeq \frac{2 t^1}{d} \left(1,0 \right) , \quad \vec \xi_{\bm 2} \simeq -\frac{2(t^1)^d}{\partial_{t^2}(Q |z^\mu|^2)} \left(0,1 \right) .
\ee
\begin{equation}
    ||\vec{\xi}_{\bm 2}||^2 = 8(t^1)^d\frac{\partial_{T^2}\partial_{\bar{T}^2}\left(Q |z^\mu|^2\right)}{\left[\partial_{t^2}(Q |z^\mu|^2) \right]^2} + \CO((t^1)^{d-1},|z^\mu|^{-2})\, .
\end{equation}
Notice that at most, derivatives of $Q$ scale like $Q$ itself or are subleading in the limit. In this case, one gets that $||\vec{\xi}_{\bm 2}||$ diverges exponentially in $|z^\mu|^{-2}$ and polynomially in $t^1$. This matches the discussion above and in \cite{Castellano:2026bnx}, where metric essential instantons are associated with strings with diverging charge vectors, signalling gravity decoupling. Furthermore, one can compute the product between extremal and rigid strings, getting to 
\begin{equation}
    \vec \xi_{\bm 1} \cdot \vec \xi_{\bm 2} \lesssim t^1 \ ,
\end{equation}
reflecting the discussion around \eqref{generalbound}. Thus, the rigid sector becomes orthogonal with respect to the SDC one.

Let us now switch to the charge vectors for BPS particles. Parametrising their central charge as $Z_\mathsf{q} = e^{K/2} (T^1)^{n_1} (T^2)^{n_2} (z^\mu)^{n_3}$, their norm reads
\be
||\vec \zeta_{\mathsf{q}}||^2 \simeq \frac{4\pi (r_2^{\mu})^2 n_3^2 (t^1)^d}{\partial_{t^2} \partial_{t^2}(Q |z^\mu|^2)}\,,\quad ||\vec \zeta_{\mathsf{q}}||^2_{n_2=n_3=0} \simeq \mathcal{O}(1).
\ee
To analyse this behaviour, we contrast the general expression above with a simplified extremal configuration in which $n_2 = n_3 = 0$. Their products with the charge vectors of the strings are
\be
\vec \xi_{\bf 1} \cdot \vec \zeta_{\mathsf{q}} \simeq \mathcal{O}(1)\,,\quad \vec \xi_{\bm 2} \cdot \vec \zeta_{\mathsf{q}} \simeq -\frac{2\pi r_2^{\mu} n_3 (t^1)^d}{\partial_{t^2}\partial_{t^2}(Q |z^\mu|^2)}
\ee
while in the extremal case, we have
\be
\vec \xi_{\bf 2} \cdot \vec \zeta_{\mathsf{q}}|_{n_2=n_3=0} \simeq \mathcal{O}(1)\,.
\ee
One can also compute their products with $\vec \zeta_*$. For generic values of the exponents, one has
\begin{equation}
\begin{split}
    &\vec \zeta_* \cdot \vec \zeta_{\mathsf{q}} \simeq n_3\frac{r_1^\mu t^1}{2\pi} + n_3 \frac{r_2^\mu t^1}{2\pi} \frac{\partial_{t^1}\partial_{t^2}(Q |z^\mu|^2)}{\partial_{t^2}\partial_{t^2}(Q |z^\mu|^2)} \lesssim t^1 \ .
\end{split}
\end{equation}
while for the exponents which identify particles in the extremal sector:
\begin{equation}
\begin{split}
    &\vec \zeta_* \cdot \vec \zeta_{\mathsf{q}} \stackrel{n_2=n_3=0}{\simeq} \CO(1) \ .
\end{split}
\end{equation}
Altogether, the charge vectors of BPS particles also organise themselves into orthogonal sectors. 

The computation changes significantly when $B\neq 0$, since the axionic shift symmetry is no longer preserved, and it is broken by a term which is parametrically larger than the would-be string tension. Consequently, the formulas for the axionic string tensions cannot be evaluated using \eqref{Tstring}. If one nevertheless did, one may check that then $\vec{\xi}_{\bm e}\cdot\vec{\zeta}_{\star}\neq 1$, a feature that clearly distinguishes them from a conventional axionic string.

\section{Laplacian and curvature}

As stressed in \cite{Etheredge:2023zjk,Aoufia:2025ppe}, Laplacians of charge vectors and other physical operators contain non-trivial information about asymptotic limits in moduli space. In the presence of axionic shift symmetries, one may evaluate their action on string tensions. For this, we consider the identity
\be
\frac{\Delta F}{F} = ||\vec{\nabla} \log F||^2 + \Delta \log F,
\label{Lap1}
\ee
for $F$ any real function. Setting $F = \CT_{\bm e}$ one obtains
 \be
\frac{\Delta \CT_{\bm e}}{\CT_{\bm e}} = \gamma_{\bm e}^2 + \Delta \log \CT_{\bm e} \lesssim \gamma_{\bm e}^2 ,
\ee
where we have used that the $\log$ is a concave function. To compute $ \Delta\CT_{\bm e}$ one may use the following identity
\be
[\Delta, \vec{\nabla}] F = {\bm R} \, \vec{\nabla} F \implies \Delta \vec{\nabla} K = {\bm R}\, \vec{\nabla} K,
\label{Lap2}
\ee
where ${\bm R}^i{}_j \equiv g^{ik} R_{kj}$ is the Ricci operator, and we have used that $ \Delta K = 2n_V$. From here we find
\be
\Delta \CT_{\bm e} = - \oh \vec{e} \cdot {\bm R}\, \vec{\nabla} K .
\label{DeltaT}
\ee
Hence, if sufficient axionic symmetries are present, $\Delta \CT_{\bm e}$ can be expressed as a linear combination of string tensions. If one now estimates the scalar curvature $R$ as the largest eigenvalue of the Ricci operator, with eigenvector $\vec{e}_{\rm max} = ({\bm e}_{\rm max}, {\bm 0})^T$, one finds that
\be
R \lesssim \gamma^2_{{\bm e}_{\rm max}}.
\label{finalR}
\ee
Hence, a scalar curvature divergence leads to a string charge-to-tension divergence and to a rigid limit \eqref{RFTmetric}, in agreement with the Curvature Criterion \cite{Marchesano:2023thx}. This is also reminiscent of the bound $R \lesssim \gamma^2_{\mathsf{q}, {\rm max}}$ observed in \cite{Castellano:2024gwi}, where $\gamma_{\mathsf{q}, {\rm max}}$ is the BPS particle charge-to-mass ratio with the largest asymptotic divergence. In that case, and for limits where $R \to \infty$, $\gamma_{\mathsf{q}, {\rm max}}$ was identified with the core RFT sector. From the current perspective, since the normalised Pauli couplings of the core RFT are the ones that dominate the scalar curvature \cite{Castellano:2026bnx}, one expects that ${\bm e}_{\rm max}$ specifies the core RFT sector of strings. 

\section{Conclusions}

\begin{figure}[htbp]
    \centering
\hspace{-0.12\linewidth}\includegraphics[width=\columnwidth]{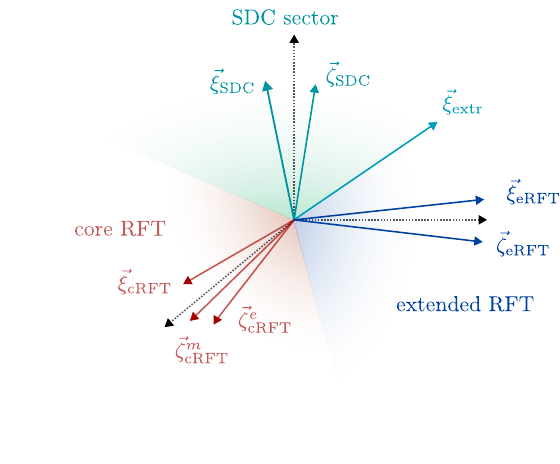}
    \vspace{-.1\linewidth}
    \caption{The three EFT subsectors (SDC, extended and core RFT) emerging along rigid limits. The charge vectors of each subset become asymptotically orthogonal to the other two.}
    \label{fig:vectors}
\end{figure}

In this work we have studied the relation between axionic shift symmetries, charge vectors and gravity decoupling. The main role of axionic shift symmetries is to identify a set of BPS axionic strings, which fix a preferred frame for the K\"ahler potential via \eqref{Tstring} along the whole asymptotic region of gravity decoupling, in agreement with general expectations \cite{Komargodski:2010rb,Adams:2011vw}. Additionally, these axionic strings define their own charge-to-tension vector fields $\vec{\xi}_{\bm e}$ which, together with the charge-to-mass vectors of BPS particles $\vec{\zeta}_{\mathsf{q}}$, map the different subsectors of the EFT into subspaces of the tangent field space. More precisely, if one splits the EFT into subsectors with different physical properties with respect to the gravity decoupling limit, they map into orthogonal subspaces. The result is a general, geometric EFT description of gravity decoupling, illustrated in figure \ref{fig:vectors}.

To arrive at this picture, we have focused on the vector multiplet moduli space of 4d $\CN=2$ EFTs that arise from type II Calabi--Yau compactifications, which have been recently studied in \cite{Marchesano:2023thx,Marchesano:2024tod,Castellano:2024gwi,Castellano:2026bnx} to revisit the process of gravity decoupling from a Swampland perspective. It follows from these works that one can distinguish between different EFT subsectors. The first subsector corresponds to extremal strings and particles, whose charge vectors have an $\CO(1)$ norm. These translate respectively into field theory directions that see a genuine supergravity metric, like for instance the leading direction ${\bm e}_0$ of the limit \eqref{growth}, and to those quantised U(1)s sourced by extremal BPS particles, like the ones within the SDC tower. These sectors are represented by charge vectors whose relative angles $\vartheta$ satisfy $\cos \vartheta \simeq \CO(1)$, as implied by certain lower bounds on their vector products. 

The second subsector corresponds to BPS strings and particles that are rigid, in the sense that their ratio $\gamma$ is parametrically above the extremality bound. Rigid axionic strings indicate field theory directions that satisfy \eqref{RFTmetric} and therefore see a rigid metric, while non-extremal BPS particles source U(1)s with rigid mutual couplings. By deriving upper bounds on the products of these vectors, we have shown that they become orthogonal to the charge vectors of certain extremal objects related to the SDC tower, that we have dubbed the SDC sector. This geometric feature corresponds to a necessary condition for gravity decoupling, since two orthogonal string vectors describe two field theory directions with negligible kinetic mixing, while orthogonal particle vectors signal the absence of gauge kinetic mixing. 

Further orthogonality relations are revealed upon inspection of the rigid sector. The extended rigid sector consists of vector multiplets that decouple from the SDC sector at the level of kinetic mixing, but not necessarily at the level of Pauli couplings \cite{Castellano:2026bnx}. Within this sector, one can identify a distinguished subsector, the core RFT, which can be characterised in terms of the monodromy of the limit and can fully decouple from the rest of the EFT. By definition, the core RFT contains electromagnetic pairs of particles \cite{Castellano:2026bnx}. We have found that, in order for their electric and magnetic U(1) couplings to reproduce the standard field theory inverse relations, the charge vectors of these electro-magnetic pairs need to align. The orthogonality of the core RFT with respect to other extended RFT charge vectors stems from the parametrically larger norms, which result in the charge-to-tension ratios $\gamma_{\bm e}$, $\gamma_{\mathsf{q}}$ with largest divergence. These results relate the notions of co-scaling and alignment put forward in \cite{Reece:2025zva} (see also \cite{Blanco:2025qom}), although here these properties are not necessarily linked to towers of particles, but to rigid sectors closed under electro-magnetic duality. Finally, by considering the Laplacian of axionic string tensions, we have seen how string ratios bound the scalar curvature, similarly to the relation found in \cite{Castellano:2024gwi}, taking us one step closer to understanding the physics behind the Curvature Criterion \cite{Marchesano:2023thx}. 

While we have focused our analysis on 4d $\CN=2$ settings, one can easily argue that our results extend to gravity decoupling limits within 4d $\CN=1$ EFTs. Indeed, the main ingredient behind our analysis is the structure \eqref{periodv}, that stems from the presence of asymptotic axionic shift symmetries and their monodromies. Such a structure has already been highlighted and exploited in the context of type II/F-theory Calabi--Yau compactifications with fluxes, see e.g. \cite{Herraez:2018vae,Escobar:2018tiu,Escobar:2018rna,Grimm:2019ixq,Marchesano:2021gyv,Bastian:2021hpc}. In this 4d $\CN=1$ setting one should interpret the period vector entries as BPS membrane tensions, instead of particle masses. Due to their no-force condition, membrane charge vectors will have similar properties to the particle charge vectors in our discussion. In particular, the tension of the lightest extremal membrane will correspond to the charge vector $\vec{\zeta}_*$ in \eqref{mstar} which, just like the BPS axionic string tensions, fixes a preferred K\"ahler frame. Identities like \eqref{stringstar} and other bounds on vector products will hold,\footnote{In this more general setup the lightest EFT string tension cannot be identified with $\Lambda_{\rm sp}^2$, but only with an upper bound for it \cite{Martucci:2024trp}.} and from there the orthogonality structure illustrated in figure \ref{fig:vectors} will follow. Since now we identify the $\vec{\zeta}$ with membrane charge vectors, the orthogonality relations encode how rigid field theory potentials embed into supergravity. This yields a new framework to understand how scalar potentials behave in asymptotic regions of field space, on which we plan to report in the future.

\section*{Acknowledgements}
We would like to thank Alberto Castellano, Damian van de Heisteeg, Luca Martucci and Lorenzo Paoloni for helpful discussions. C.A. would like to thank the II. Institute for Theoretical Physics at the University of Hamburg and the DESY Theory Group 
for hospitality during the different stages of this work. This work is supported through the grants CEX2020-001007-S, PID2021-123017NB-I00 and PID2024-156043NB-I00, funded by MCIN/AEI/10.13039/501100011033, ERDF, EU.	

\bibliography{papers}
\end{document}